%!TEX TS-program = pdftex

\message{ Add fonts 
  Roman & Sans }
\font\titolo=cmss17 
\font\smallcaps=cmcsc10 

\def\A{{\cal A}}
\def\Rfi{{I\mskip-7mu R}}
\def\R{{I\mskip-7mu R}}

\magnification1200

{\voffset=6.8truecm
\hoffset=1.2truecm
\hsize=14truecm
\vsize=19truecm
%\nopagenumbers
\noindent Appeared in: {\it   Differential and Integral Equations 4}, 305--329 
(1991).
\bigskip\bigskip 
{\baselineskip12truept

\centerline{{\titolo A Class of 
    Integrable Hamiltonian Systems}}

\centerline{{\titolo Including Scattering of
                     Particles on the Line}}  

\centerline{{\titolo with Repulsive
                     Interactions}} 
}

\medskip\smallskip                                                            

\centerline{{\smallcaps Gianluca Gorni}}
\centerline{{\it Universit\`a di Udine}}
\centerline{{\it Dipartimento di Matematica 
             e Informatica}}
\centerline{{\it
             via Zanon 6,
             33100  Udine, Italy}}
\medskip
\centerline{{\smallcaps Gaetano Zampieri}}
\centerline{{\it Universit\`a di Padova}}
\centerline{{\it Dipartimento  di Matematica
        Pura e Applicata}}
\centerline{{\it
              via   Belzoni 7, 
              35131 Padova, Italy}}

\medskip\vfil

%\centerline{August, 1989}

\bigskip
 
{\bf Abstract.}
The main purpose of this paper is to introduce a
new class of Hamiltonian scattering systems of
the cone potential type that can be integrated via
the asymptotic velocity.  For a large subclass,
the asymptotic data of the trajectories define a
global canonical diffeomorphism $\A$ that brings
the system into the normal form $\dot P=0$,  $\dot
Q=P$.

The integrability theory applies for example to
a system of $n$ particles on the line interacting
pairwise through rather general repulsive
potentials. The inverse $r$-power
potential for arbitrary~$r>0$ is included, 
the reduction to normal form being carried out
for the exponents~$r>1$. In particular, the
Calogero system is obtained for~$r=2$. The
treatment covers also the  nonperiodic
Toda lattice.

The cone potentials that we allow
can undergo small perturbations in any arbitrary 
compact set without losing the integrability and
the reduction to normal form.

\vfil

\centerline{This research was supported by the
     {\it Ministero della Pubblica Istruzione}
            and by the C.N.R.}

\eject
}

\font\bfuno=cmbx12
\font\smallcaps=cmcsc10
\font\sf=cmss10

\def\V{{\cal V}}
\def\C{{\cal C}} 
\def\D{{\cal D}}
\def\A{{\cal A}}
\def\M{{\cal M}}
\def\sss{\scriptscriptstyle}

%%%%%%%%%%  AMS  fonts!  %%%%%%%%%%%%%%%
\input amssym.def
\input amssym.tex

%%%%%%%%%%%%%%%%%%%%%%%%%%%%%%%%%%%%%%%%%

\def\mathbb{\Bbb}
\def\R{{\Bbb R}}
\def\Rfi{\R}

\def\Rn{{\R^n}}

\def\Dif{\,\hbox{\sf D}}
\def\Rn{{\R^n}}
\def\pinf{{p_{\sss\infty}}}
\def\ainf{{a_{\sss\infty}}}
\def\pqbar{{\bar p,\bar q}}
\def\pqzero{{\bar p_{\sss0},\bar q_{\sss0}}}
\def\dom{{\hbox{\rm dom}\,}}
\def\dist{{\,\hbox{\rm dist}}}
\mathchardef\emptyset="083F
\def\qed{%
  \vbox{\hrule\hbox to7.8pt{\vrule height7pt
  \hss\vrule height7pt}\hrule} }

%%%%%%%%%%%%%%%%%
\pageno=2

\centerline{{\bfuno 1. Introduction}}
\bigskip

An authonomous $2n$-dimensional Hamiltonian
system is said to be integrable if there exist
$n$ smooth first integrals, independent and in
involution (see e.g.~[A] for details). The {\it
scattering} systems provide natural constants of
motion: the {\it asymptotic velocities}.
Unfortunately, there is no obvious reason for
them to be smooth functions of the initial data,
and in fact they are sometimes not even
continuous (see e.g.~[GZ1], Section~3). Some work
has been done to single out classes of
scattering-type systems for which rigorous proofs
of smoothness ($C^k$, $2\le k \le+\infty$) of the
asymptotic data and of integrability could be
carried~out.

In [GZ1] we investigated the {\it complete
integrability} of Hamiltonian systems
of the form 
$$\dot p=-\nabla\V(q)\,,\quad \dot q=p\,,
  \qquad p,q\in\Rn,
  \eqno(1.1)$$
where $\V$ is a $C^k$ ($3\le k\le+\infty$)
potential---defined on a domain that we
  can suppose for now to be simply $\Rn$---with
the following basic properties: 

\medskip

\item{CP$_1$} it is {\it bounded below} (say,
  $\V\ge0$);

\item{CP$_2$} it is a {\it cone potential}, which
means the following. Let $\C$ be the convex cone
in~$\Rn$ generated by the forces $-\nabla\V$
(i.e., the set of all linear combinations of
vectors of the form $-\nabla\V(q)$, with
nonnegative coefficients). Then the closure
of~$\C$ is a {\it proper cone}, that is, it
contains no straight line.

\medskip

\noindent
Condition CP$_2$ is equivalent to requiring that
the {\it dual cone}
$$\D:=\{w\in\Rn\;:\;w\cdot v\ge0\quad\forall
                    v\in\C\}
  \eqno(1.2)$$
has nonempty interior: $\D^\circ\ne\emptyset$.

The first simple consequences of CP$_1$ is that
each solution of the system~(1.1) is globally
defined in time and has a bounded velocity (just
remind that the energy $|p|^2/2 +\V(q)$ is
conserved). On the other hand, because of~CP$_2$,
$\D$ contains a basis $b_1,\ldots,b_n$ of~$\Rn$.
The scalar product $p(t)\cdot b_i$ of the
velocity of a trajectory with an element of the
basis is monotonic in time (the derivative is
$-\nabla\V(q(t))\cdot b_i$, which is nonnegative
for how $\D$ is defined).
Altogether, we get the existence of the
following finite limit:
$$\pinf(\pqbar)=\lim_{t\to+\infty}
  p(t,\pqbar)\in\Rn\,,
  \eqno(1.3)$$
the {\it asymptotic velocity}. We denote by
$t\mapsto(p(t,\pqbar), \allowbreak q(t,\pqbar))$
the motion of the system~(1.1) which has
$(\pqbar)$ as initial
conditions.

Of course, $\pinf$ is a constant of motion.
In~[GZ1] we found sufficient conditions for
the components of~$\pinf$ to be smooth ($C^k$,
$2\le k\le+\infty$), independent and pairwise in
involution first integrals of motion. This led
to a large class of completely integrable
systems of the form~(1.1).

\goodbreak

In the course of our investigations we figured
out the graph of an ideal cone potential as a
sort of smooth, convex ``amphitheatre'' surface,
with hyperbolic-type level sets each asymptotic
to the boundary of a translation of~$\D$; the
surface flattens down to zero in the
``direction'' of the dual cone~$\D$ and steeps up
in the opposite directions. Part of the job was
imagining what pathologies may arise in the mere
hypotheses a), b) and~c), and finding further
conditions that could rule them out.

\goodbreak

The problem of the regularity of~$\pinf$ was
attacked through the following relation:
$$\pinf(\pqbar)=\bar p+\int_0^{+\infty}
  \mskip-14mu-\nabla\V(q(s,\pqbar))\,ds\,,
  \eqno(1.4)$$
which is an easy consequence of~(1.1) and~(1.3).
By formally differentiating~(1.4) $n$~times with
respect to the initial data $(\pqbar)$, we
obtain formulas containing the differentials
of~$\V$, evaluated always in $q(s,\pqbar)$, and
the derivatives of $q(s,\pqbar)$ again with
respect to~$(\pqbar)$.

If we wish to apply the theorems on the regular
dependence of integrals from parameters, we seem
to need the following:

\medskip

\item{1} information on how the
trajectory $q(t,\pqbar)$ behaves as $t\to+\infty$,
somehow locally independent of the initial data;

\item{2} bounds on the growth of the derivatives
of $q(t,\pqbar)$ with respect to~$(\pqbar)$,
as~$t\to+\infty$.

\medskip

\noindent
The decay rate of~$\V$ along the trajectories can
then be chosen so as to compensate for the
growth  of the derivatives of $q(t,\pqbar)$ and
yield the desired properties for formula~(1.4)
and the derived ones.

Point 1 is what the results of the present paper
are all about. The techniques developed in~[GZ1]
to cope with problem~2 apply with no change to
the new situations, and we are not going to dwell
on them for longer than the following remarks.
Within hypotheses of {\it exponential decay} of
the derivatives of~$\V$ in the direction of the
dual cone~$\D$, we could use a Gronwall argument
to obtain an a~priori less-than-exponential
bound on the growth of the derivatives of
$q(t,\pqbar)$ with respect to the initial data
as~$t\to+\infty$. Otherwise,
much more general decays (such as
inverse $r$-power for arbitrary~$r>0$) could be
allowed if {\it convexity} and a kind of
{\it monotonicity} in the Hessian matrix of~$\V$
were imposed, in order to exploit certain
Liapunov functions for the first variational
equations.

In [GZ2] we somewhat specialize our
hypotheses (the ones connected with point~2 only)
to permit the reduction of~(1.1) to the {\it
normal form} 
$$\dot P=0\,,\quad \dot Q=P\,,
  \eqno(1.5)$$
by means of the global canonical
transformation $(P,Q)=\A(p,q)$ ({\it asymptotic
map}) defined as
$$\A\colon(\pqbar)\mapsto\bigl
  (\pinf(\pqbar),\ainf(\pqbar)\bigr)\,,
  \eqno(1.6)$$
where $\ainf(\pqbar)$, the {\it asymptotic
phase}, is given by
$$\ainf(\pqbar):=\lim_{t\to+\infty}
  \bigl(q(t,\pqbar)-t\,\pinf(\pqbar)\bigr)\,,
  \eqno(1.7)$$
(this limit always exists within the assumptions
of~[GZ2]). 

In the framework of [GZ1], the complete
integrability has a certain property of
``structural stability'', in the precise sense
that it is persistent under small perturbations
of the potential~$\V$ in an arbitrary compact set
of~$\Rn$. In~[GZ2] the map $\A$ enjoys the same
property.

\medskip
\line{\hfil * \qquad * \qquad * \hfil }
\goodbreak
\medskip

The precise nature of the information we refer
to in point~1 is the following:

\medskip

\item{AR} we have global {\it asymptotic
  regularity}, that is, the asymptotic velocity
  $\pinf$ belongs to the {\it interior} of the
  dual cone~$\D$ for {\it all} initial data:
  $\pinf(\pqbar)\in\D^\circ$
  $\forall(\pqbar)$; this ensures in particular
  that, for every single trajectory, the distance
  of $q(t,\pqbar)$ from the boundary~$\partial\D$ 
  of the dual cone~$\D$ grows linearly
  as~$t\to+\infty$;

\item{UE} we have {\it estimates} from below
  on the growth of
  $\dist(q(t,\pqbar),\partial\D)$ that are 
  {\it locally uniform} in the initial data.
  More precisely, this distance is not less than
  $\gamma t$ for all times~$t$ larger
  than~$t_{\sss0}$, where $\gamma>0$ and
  $t_{\sss0}\in\R$ are locally independent
  of~$(\pqbar)$.

\medskip

\noindent
The basic reason why we use the quantity 
$\dist(q,\partial\D)$ is that it has a very
simple analytic expression (see Section~2
of~[GZ1] and Proposition~2.2 of the present
paper), in terms of relevant quantities of
our problem. For example, the decay rate of~$\V$
``at~infinity'' is naturally written in terms of
this distance for all the examples that
motivated our study. 

\goodbreak

In [GZ1] the two points AR and~UE were treated
in Section~4, Propositions~4.3 and~4.4
respectively, and they were tightly intertwined.

The result of Section~3 of this paper is
the proof that, under mild conditions on~$\V$ 
(not implied by~[GZ1], Section~4), the two
problems can be treated independently
of each other. Namely, locally uniform estimates
arise naturally in the invariant set~$\M$ of the
initial data which are {\it asymptotically
regular}, that is, whose trajectories have
asymptotic velocity in the interior of the dual
cone: $$\M:=\{(\pqbar)\;:\;
  \pinf(\pqbar)\in\D^\circ\}\,.
  \eqno(1.8)$$
The set $\M$ turns out to be open and nonempty. 
The locally uniform estimates that we derive
permit to apply the smoothness theory of~[GZ1]
and~[GZ2] and obtain complete integrability
and normal form for ``restricted phase space''
systems 
$$\dot p=-\nabla\V\,,\quad \dot q=p\,,
  \qquad (p,q)\in\M\,.
  \eqno(1.9)$$
Actually, the methods of Section~3 seem to be
general enough to be possibly adapted to
Hamiltonian systems where only subsets of the
phase space are of cone potential type. For each
of these subsets we can single out further
subsets on which the system is integrable
via asymptotic velocity.

\goodbreak

One may wonder whether a cone potential system
may be globally integrated via asymptotic
velocity even if there is no global asymptotic
regularity. The answer is ``no'' for all
systems for which the range of~$\pinf$
is contained in the closed set~$\D$ (this
property is an easy consequence of quite natural
assumptions on the potential; see  e.g.\
Proposition~4.10). 
In fact, if $\pinf$ takes some values on the
boundary of~$\D$, then it is not an open map. In
particular, even if $\pinf$ were globally
smooth, its components could not be independent
at those points where $\pinf\in\partial\D$.

Of course, a draw-back of the results of
Section~3 is that the invariant set~$\M$, on
which we establish integrability, is defined
{\it implicitly}, through asymptotic properties
of the trajectories. It is hard in general to
decide whether given initial data are
asymptotically regular or not. The negative
example of Section~5 goes the other way
round, that is, it starts with a single
trajectory with a suitable behaviour and builds
upon it a potential that admits that trajectory
as a non asymptotically regular
solution. 
However, we already know
from~[GZ1], Section~4, a wide class of cone
potentials for which $\M$ coincides with the whole
configuration space. In Sections~6 and~7 of the
present paper we will describe two new classes
with that property, coming up with new {\it
globally integrable systems}. These two classes
are distinct from each other and from the one
in~[GZ1], although there are mutual
overlappings.

The major shortcoming of the conditions stated
in [GZ1], Section~4, is that they
allow only cones $\C$ of width not larger
than~$\pi/2$:
$$\nabla\V(q^\prime)\cdot
  \nabla\V(q^{\prime\prime})>0\quad
  \forall q^\prime,q^{\prime\prime}.
  \eqno(1.10)$$
This restriction ruled out for example the
well-known non-periodic Toda lattice system, the
Calogero system and, more generally, the
problem of $n$~particles on the line with
pairwise repulsive interactions, that had
actually been the original motivation to
introduce the concept of cone potential.
The two new classes whose integrability we are
going to prove here cover many
``wide cone'' systems.

\goodbreak

We start in Section 4 with restricting our
attention to the potentials with the particular
{\it finite-sum} form:
$$\V(q):=\sum_{\alpha=1}^N
  f_\alpha(q\cdot v_\alpha)
  \eqno(1.11)$$
($N\ge1$, no relation to $n$), where the
$v_\alpha$ are vectors of~$\Rn$ and the
$f_\alpha$ are smooth real functions of one
variable. Since the force $-\nabla\V$ is
$$-\nabla\V(q)=\sum_{\alpha=1}^N-f^\prime_\alpha
  (q\cdot v_\alpha)v_\alpha\,,
  \eqno(1.12)$$
$\V$ is a bounded below cone
potential whenever the functions $f_\alpha$ are 
bounded below and all (say) decreasing, and the
convex cone generated by the vectors $v_\alpha$
is proper (this cone obviously contains the cone
of the forces~$\C$, and coincides with the
closure~$\bar\C$ under simple conditions
on the functions~$f_\alpha$, see
Proposition~4.2). 

\goodbreak

The potential of a system of
$n$~particles on the line interacting pairwise
through the repulsive potentials $f_{ij}$ is
given by  
$$\V(q_1,\ldots,q_n)=\sum_{1\le i<j\le n}
  f_{ij}(q_i-q_j)\,,
  \eqno(1.13)$$
and can be written in the form~(1.11) by
defining the vectors
$$v_{ij}:=(\ldots,0,{\buildrel(i)\over1},
  0,\ldots,0,{\buildrel(j)\over{-1}},0,\ldots)\,,
  \quad 1\le i<j\le n\,.
  \eqno(1.14)$$
The cone generated by the $v_{ij}$ is
wider than~$\pi/2$, since for example
$$v_{1,2}\cdot v_{2,3}=(1,-1,0,0,\ldots)\cdot
  (0,1,-1,0,\ldots)=-1
  \eqno(1.15)$$
and these two vectors form an angle of~$2\pi/3$.
The dual cone is
$$\{(q_1,q_2,\ldots,q_n)\in\Rn\;:\;
  q_1\le q_2\le\ldots\le q_n\}\,,
  \eqno(1.16)$$
which has nonempty interior.

\goodbreak

To be able to deal with wide cones we make heavy
use of the finite-sum form of~$\V$. We
emphasize that the present results are distinct
from the ones in~[GZ1] and~[GZ2], that
allowed both finite-sum potentials that do not
fit into the new classes and examples of a
different, non finite-sum structure.

In Section 4 we establish sufficient conditions
on the finite-sum potentials to give rise to
``restricted phase space'' integrability. Of
course, the main conclusions rest upon the
theory develped in~[GZ1] and~[GZ2]. We also
provide a number of lemmas meant for the sequel. 

In Section~5, as we already mentioned, we provide
an example of a finite-sum potential admitting
non asymptotically regular initial data.
The counterexamples already given in~[GZ1],
Section~3, were, in a way or another, related to
the presence of equilibria (even if not strictly
speaking). The reason for the failed
asymptotic regularity in the present case seems
to be subtler. Figuratively speaking, there are
two forces acting on~$\R^2$, that largely oppose
each other, because the angle between them
is~$2\pi/3$. One of the two decreases rapidly
(exponentially) at infinity, while the other
decays slowly (the potential is not integrable at
infinity, behaving as $x\mapsto1/x$). It is
proved in Section~3 that all trajectories that
start with a large enough speed, pointing into
the interior of the dual cone~$\D$, are
asymptotically regular, because the particle
escapes so fast that the ever dwindling forces
never catch up to drive the velocity toward the
boundary of~$\D$. However, it happens that if the
initial speed is too small, then the
slow-decaying potential takes over the motion,
the distance of the particle from the boundary of
$q_{\sss0}+\D$ (all trajectories are contained in
a set of this form) grows logarithmically, and
the asymptotic velocity belongs to the boundary
of~$\D$.

In Section~6 we prove that if the functions
$f_\alpha$ are {\it all equal}, or, at least,
multiples of a single function:
$f_\alpha=c_\alpha f$, $c_\alpha>0$,
plus a nonrestrictive technical assumption
on~$f$, then all initial data are asymptotically
regular. Roughly speaking, none of the single
forces can prevail, because they  all decay
the same way. The global complete integrability
holds for all potentials in this class, whilst
the reduction into normal form via asymptotic
map is provided only if $f$ is integrable, case
which intersects with the next class.

\goodbreak

In Section~7 we show that global asymptotic
regularity holds if the functions $f_\alpha$ are
{\it all integrable} at~$+\infty$. In the
language of the counterexample, the forces are
too weak at infinity to bend enough the
trajectories. For all the potentials in this
class the asymptotic map is defined and smooth.
The proof in this Section~7 is adapted
from~[Gu4], where the author also admits somewhat
more general potentials than here, but does not
address the problem of the smoothness of the
asymptotic map (he calls ``regular'' the
trajectories which are asymptotically regular in
our sense and possess asymptotic phase).

The class of finite-sum potentials with global
asymptotic regularity that was described
in~[GZ1], Section~10, corresponds to no
special decay or similarity conditions on the
functions~$f_\alpha$, but required instead that
{the cone $\C$ were not wider than~$\pi/2$}
($v_\alpha\cdot v_\beta\ge0$
$\forall\alpha,\beta$). There the forces
actually agree so much in direction that there
is not enough competition.

Of course, any future attempt to find further
classes of cone potentials with global
asymptotic regularity must take into account the
existence of counterexamples. Actually, there
appears to be not much room for extensions
within the finite-sum form.

\medskip
\line{\hfil * \qquad * \qquad * \hfil }
\goodbreak
\medskip

In Section~8 we gather into one statement all
the hypotheses on finite-sum potentials under
which we will have proved global integrability
via asymptotic velocity. We are now going to
write down some explicit, remarkable examples
with $C^\infty$-integrability. In these
examples $N\ge1$ and the vectors 
$v_\alpha\in\Rn$, with $\alpha=1,\ldots,N$,
generate a proper cone (i.e., there exists $\bar
v\in\Rn$ such that $\bar v\cdot v_\alpha>0\;\;
\forall\alpha$).

\smallskip

\item{a)} Toda-like, or exponential potentials
$$\V(q):=\sum_{\alpha=1}^N c_\alpha 
  e^{-q\cdot v_\alpha}\,,\qquad
  c_{\alpha}>0,\quad q\in \Rfi^n\,.
  \eqno(1.17)$$

\item{b)} Finite sum of inverse powers:
$$\V(q):=\sum_{\alpha=1}^N
  {1\over(q\cdot v_\alpha)^{r_{\alpha}}}\,,\qquad
  q\in \{\bar q\in\Rfi^n\;\colon\;
  \bar q\cdot v_\alpha>0\;\forall\alpha\}\,,
  \eqno(1.18)$$
in three different hypotheses:

\medskip

\itemitem{b1)} $r_\alpha>0$ arbitrary but the
vectors verify $v_\alpha\cdot v_\beta\ge0$;

\itemitem{b2)} $v_\alpha$ arbitrary but all the
exponents $r_\alpha$ are~$>1$;

\itemitem{b3)} $v_\alpha$ arbitrary but the
exponents $r_\alpha$ are all equal to an~$r>0$.

\medskip

\item{}
In particular b3) includes the inverse $r$-power
potential for arbitrary $r>0$:
$$\V(q):=\sum_{1\le i<j\le n}
  {1\over(q_i -q_j)^r}\,,\qquad q\in
  \{(q_1,\ldots,q_n)\in\Rfi^n\,:\,
  q_1<q_2<\ldots<q_n \}\,.
  \eqno(1.19)$$
The case $r=1$ is the Coulombian potential,
whilst $r=2$ is the Calogero potential. The
exponent $r=1$ divides what are usually called
short range potentials ($r>1$) from the long
range ones ($0<r\le1$). 

\medskip

\item{c)} Even longer range potentials are the
inverse-logarithmic type:
$$\V(q):=\sum_{\alpha=1}^N
  {1\over(\ln(1+q\cdot v_\alpha))^r}\,,\qquad
  q\in \{\bar q\in\Rfi^n\;\colon\;
  \bar q\cdot v_\alpha>0\;\forall\alpha\}\,,
  \eqno(1.20)$$
for arbitrary $r>0$.

\medskip

For potentials a), b1) with $r_\alpha>1$, and b2),
there are both asymptotic velocities and phases,
with consequent reduction to normal form. In
case~b3) with $0<r\le1$ and c) we   prove
integrability through asymptotic velocities but
we do not have the  reduction to normal form.

\medskip
\line{\hfil * \qquad * \qquad * \hfil }
\goodbreak
\medskip

The earliest cone potential systems to be
integrated, with a complete
description of the motions, were some systems of
particles on the line~$\R^1$: most notably the
non-periodic Toda lattice ({\it exponential}
potentials) and the Calogero system ({\it
inverse square} potential). The methods  used in
the proof were, most generally, Lax pairs and
isospectral deformations (see  e.g.~[M]). As a
side product, asymptotic velocities turned out to
be a set of analytic, independent and pairwise in
involution first integrals.

Gutkin in~[Gu1] introduced the concept of cone
potentials with the conjecture that the
integrability of those special systems could be
derived from a general theory.
Gutkin carried on the study in
some subsequent papers ([Gu2 to~4]). However,
at the beginning of the present research, the
earliest article we were aware of with a rigorous
proof of $C^\infty$-integrability via asymptotic
velocities for cone potentials was the
recent~[OC], where Oliva and Castilla studied
mainly finite-sum potentials for which the
$f_\alpha$ decay exponentially at~$+\infty$ and
the vectors $v_\alpha$ generate a cone not wider
than~$\pi/2$, and a few special cases with wider
cones. Their proof used strongly the finite-sum
form of the potential to define a
``compactifying'' change of variable, and then
applied a Lemma in Dynamical Systems on the
differentiability of a foliation of invariant
manifolds (developed and proved in the same
paper, but of independent interest as well).

When we had already completed the bulk of the
present work, we came to know Hubacher's
paper~[Hu], whose results overlap partially with
ours. It is concerned with systems
of mutually repulsive particles on the line, that
is, with potentials of the form~(1.13). To start
with, it gives results of what we call global
asymptotic regularity---see~AR above---analogous
to the ones in Section~6 and~7 of the present
paper, with different proofs, tailored to the
structure of the mutually repelling particles on
the line. It also provides a counterexample
to global asymptotic regularity that reminds
ours: the potential is the sum of exponentials
and an inverse power (the latter is however
$x\mapsto x^{-\alpha}$ with $\alpha=1/2$, and not
$-1$ as here; $-1$~is a limit exponent, because
the case $\alpha>1$  is short-range and global
regularity holds, see Section~7). Then Hubacher
restricts her attention to the systems with short
range repulsive equal potentials  
$$\eqalign{&
  \V(q_1,\ldots,q_n)=\sum_{1\le i<j\le n}
  f(q_i-q_j)\,,\cr
  &|f^\prime(x)|\le{M\over x^{2+\alpha}}\,,\quad
  |f^{\prime\prime}(x)|\le{M\over x^{3+\alpha}}
  \,,\quad M>0,\; \alpha>0\,,\cr}
  \eqno(1.21)$$
and states their integrability through
asymptotic data by invoking the results of
Simon~[S] and Herbst~[He]. Those authors had
studied the scattering of $n$ mutually repulsive
particles in~$\Rn$, through an approach totally
different from ours. The proof of smoothness
was there obtained by solving a kind of Cauchy
problem at infinity with the asymptotic data
playing the role of initial data. The method
is remarkably simple for short range potentials,
but it gets into complications for long
range potentials $x\mapsto 1/x^\alpha$ when
$0<\alpha\le1$, and does not seem to cover for
example the inverse logarithm case~(1.20).
  
Let us also say that Hubacher does not adopt the 
mere existence of $n$ independent first integrals
in involution as definition of integrability. 
Among the systems with potential~(1.21), she
prefers to reserve the name ``integrable'' to
those which satisfy  certain conditions including
a suitable preservation property between the
asymptotic velocities of the particles as
$t\to-\infty$ and the ones as $t\to+\infty$.  For
these integrable systems she obtains an
interesting result (Theorem~4 of [Hu]). However,
that kind of integrability is very special and
totally outside the spirit of the present paper.

Moauro, Negrini and Oliva [MNO] have recently
obtained a proof of {\it analytic} integrability
for the systems of cases~a) and~b) above. They
exploited the geometric techniques introduced
in~[OC], with crucial changes in the time
variable as in~[MN], and, strongly, the theory
of invariant manifolds for fixed points of
analytic diffeo\-morphisms. To their goal they
also used Propositions~6.2 and~7.2 of the present
paper.

Finally, we remark that the problem of the mere
existence of asymptotic velocities and phases for
scattering systems of mutually repelling
particles in~$\Rn$ was studied by various authors
in the sixties and seventies, see for instance
[Ga] and the references contained therein.

       \vfill\eject

%%%%%%%%%%%%%%%%%
\centerline{{\bfuno 2. Preliminaries}}  
\bigskip

The starting assumption on {\it cone potentials}
are as follows.

\bigskip

{\bf Hypotheses 2.1 } \sl The potential~$\V$ is
a $C^2$ real function defined on a nonempty,
open subset $\dom\V$ of~$\Rn$. Moreover:

\item{CP$_0$} for every $E\in\R$, the
closure in~$\Rn$ of the set
$\{q\in\dom\V\;:\;\allowbreak \V(q)\le E\}$
is contained in~$\dom\V$;

\item{CP$_1$} $\V$ is bounded below (it makes no
harm to assume $\V\ge0$);

\item{CP$_2$} the convex cone
$$\D:=\{w\in\Rn\;:\;-\nabla\V(q)\cdot w\ge0\;
  \forall q\in\dom\V\}
  \eqno(2.1)$$
has nonempty interior (to avoid trivialities we
can assume that $\D$ is not all of~$\Rn$, i.e.,
$\V$ is nonconstant).  \rm

\bigskip

From Hypotheses~2.1 it follows in particular
that $q\in\dom\V\Rightarrow\allowbreak
q+\D\subset\dom\V$. In fact, for all $w\in\D$,
the function $t\mapsto\V(q+tw)$ is (weakly)
decreasing, and hence
$\dom\V$ must contain all the half-line
$\{q+tw\;:\;t\ge0\}$.

From the conservation of energy it is easy to
see  that, for all initial data
$(\pqbar)\in\allowbreak\Rn\times\dom\V$, the
Hamiltonian system
$$\dot p=-\nabla\V(q)\,,\quad
  \dot q=p\,,\qquad
  p\in\Rn\,,\; q\in\dom\V
  \eqno(2.2)$$
has a unique solution
$t\mapsto(p(t,\pqbar),q(t,\pqbar))$, which is
globally defined in time, and with bounded
velocity: $|p(t,\pqbar)|^2\le|\bar p|^2+2\V(\bar
q)$. Next, $\D$, having
nonempty interior, contains a basis
$\{b_1,\ldots,b_n\}$ of~$\Rn$. We have
$$\dot p(t,\pqbar)\cdot b_i=
  -\nabla\V(q(t,\pqbar))\cdot b_i\ge0\,,
  \eqno(2.3)$$
so that each function $t\mapsto p(t,\pqbar)\cdot
b_i$ is monotone. Being also bounded, it has a
finite limit as~$t\to+\infty$. This proves that,
under Hypotheses~2.1, the {\it asymptotic
velocity} 
$$\pinf(\pqbar):=\lim_{t\to+\infty}
  p(t,\pqbar)
  \eqno(2.4)$$
exists for all initial data $(\pqbar)\in\Rn\times
\dom\V$.

\bigskip

We remind that by {\it cone} in~$\Rn$ we mean a
nonempty subset~$C$ of~$\Rn$ such that
$\lambda\ge0\Rightarrow\lambda C\subset C$.
All the cones we consider are convex. For a
cone~$C$ we define its {\it dual~cone~$C^*$} by
$$C^*:=\{v\in\Rn\;:\;v\cdot w\ge0\;\forall
  w\in C\}\,.
  \eqno(2.5)$$
Some properties of cones are listed in Section~2
of~[GZ1]. We only report here the following
statement, concerning the {\it distance} of a
point of a convex cone from the boundary. 

\bigskip

{\bf Proposition 2.2 } \sl Let $C$ be a 
convex cone in $\Rn$, not reduced to $\{0\}$,
and let $D$ be its dual.
If $w\in D$ then \rm
$$\hbox{dist}(w,\partial D)=
  \min_{{v\in\bar C\atop|v|=1}}\; w\cdot v\,.
  \eqno(2.6)$$

      \vfill\eject

%%%%%%%%%%%%%%%%%
\centerline{{\bfuno 3. Integrability on an
                       Invariant Set}}
\bigskip

We are going to show how a very simple
geometric condition on the cone potential~$\V$
guarantees that the set $\M$ of the
asymptotically regular initial data is nonempty
and open, and that locally uniform estimates
hold as needed to apply the smoothness theory
of~[GZ1] and~[GZ2].

In the sequel, $\C$ will be the convex cone
generated by the force $-\nabla\V$ of a cone
potential, and $\D$ will be the dual of~$\C$.

\bigskip

{\bf Hypothesis 3.1 } \sl There exists
$q_{\sss0}\in\dom\V$ and a nonnegative, weakly
decreasing and integrable function
$h_{\sss0}\colon[0,+\infty[\to\R$ such that \rm
$$q\in q_{\sss0}+\D\quad\Rightarrow\quad
  |\nabla\V(q)|\le h_{\sss0}\bigl(
  \hbox{{\rm dist}}
  (q,q_{\sss0}+\partial\D)\bigr)\,.
  \eqno(3.1)$$

\bigskip

We start with proving that there are
asymptotically regular initial data.

\bigskip

{\bf Proposition 3.2 } \sl Assume Hypotheses~2.1
and~3.1. Then, for any $\gamma>0$ there exists
$q_\gamma\in q_{\sss0}+\D$ such that \rm
$$\left.\matrix{\bar p\in\D^\circ\,,\hfill\cr
  \noalign{\smallskip}
  \hbox{dist}(\bar p,\partial\D)
  \ge2\gamma\,,\hfill\cr
  \noalign{\smallskip}
  \bar q\in q_\gamma+\D\,,\hfill\cr}
  \right\}
  \quad\Rightarrow\quad
  \forall t\ge0\quad
  \left\{\matrix{
  p(t,\bar p,\bar q)\in\D^\circ\,,\hfill\cr
  \noalign{\smallskip}
  \hbox{dist}\bigl(p(t,\bar p,\bar q),
  \partial\D\bigr)\ge\gamma\,,\hfill\cr
  \noalign{\smallskip}
  q(t,\bar p,\bar q)\in q_\gamma+\D\,,\hfill\cr
  \noalign{\smallskip}
  \hbox{dist}\bigl(q(t,\bar p,\bar q),q_\gamma
  +\partial\D\bigr)\ge\gamma t\,.\hfill\cr}
  \right.
  \eqno(3.2)$$

\bigskip

{\bf Proof. } Fix $\gamma>0$ and and pick
$b_\gamma>0$ such that
$$\int_0^{+\infty}\mskip-14mu
  h_{\sss0}(\gamma t+b_\gamma)\,dt<\gamma\,.
  \eqno(3.3)$$
Let $q_\gamma\in q_{\sss0}+\D$  be such
that dist$(q_\gamma,q_{\sss0}+\partial\D)\ge
b_\gamma$. Let the initial data $(\bar p,\bar
q)$ verify
$$\bar p\in\D\,,\quad
  \hbox{dist}(\bar p,\partial\D)
  \ge2\gamma\,,\quad
  \bar q\in q_\gamma+\D\,.
  \eqno(3.4)$$
For all $t\ge0$ such that
$q(t,\pqbar)\in q_\gamma+\D$ we define the
function 
$$\psi(t):=\hbox{dist}\bigl(q(t,\pqbar),
  q_\gamma+\partial\D\bigr)=
  \min_{v\in\bar\C\atop|v|=1}
  \bigl(q(t,\pqbar)-q_\gamma
  \bigr)\cdot v\,.
  \eqno(3.5)$$
Since $\bar p\in\D^\circ$, the trajectory
$q(t,\pqbar)$ belongs to $q_\gamma+\D$
for all $t\ge0$ small enough.

\noindent
For each $v\in\bar\C$, $|v|=1$, the function
$$t\mapsto\bigl(q(t,\pqbar)-q_\gamma
  \bigr)\cdot v
  \eqno(3.6)$$
is Lipschitz with constant $\sup_t|p(t,\pqbar)|
<+\infty$. Hence $\psi$ is Lipschitz too, with
the same constant (see the Appendix), and in
particular it is absolutely continuous. 

\noindent
In an interval contained in the domain
of~$\psi$, let $t_{\sss0}<t$ and write
$$q(t,\pqbar)-q(t_{\sss0},\pqbar)=
  (t-t_{\sss0})\,p(t_{\sss0},\pqbar)+
  (t-t_{\sss0})\varepsilon(t)\,,
  \eqno(3.7)$$
with $\varepsilon(t)\to0$ as $t\searrow
t_{\sss0}$. Let $v_t\in\bar\C$, $|v_t|=1$ be
such that
$$\psi(t)=\bigl(q(t,\pqbar)-q_\gamma
  \bigr)\cdot v_t\,.
  \eqno(3.8)$$
Then
$$\eqalign{\psi(t)-\psi(t_{\sss0})&{}=
  \bigl(q(t,\pqbar)-q_\gamma
  \bigr)\cdot v_t -
  \min_{v\in\bar\C\atop|v|=1}
  \bigl(q(t_{\sss0},\pqbar)-q_\gamma
  \bigr)\cdot v\ge\cr
  &{}\ge
  \bigl(q(t,\pqbar)-q_\gamma
  \bigr)\cdot v_t -
  \bigl(q(t_{\sss0},\pqbar)-q_\gamma
  \bigr)\cdot v_t=\cr
\noalign{\smallskip}
  &{}=
  \bigl(q(t,\pqbar)-q(t_{\sss0},\pqbar)
  \bigr)\cdot v_t =\cr
\noalign{\smallskip}
  &{}=
  (t-t_{\sss0})\,
  p(t_{\sss0},\pqbar)\cdot v_t+
  (t-t_{\sss0})\,\varepsilon(t)\cdot v_t \ge\cr
\noalign{\smallskip}
  &{}\ge
  (t-t_{\sss0})\min_{v\in\bar\C\atop|v|=1}
  p(t_{\sss0},\pqbar)\cdot v -
  (t-t_{\sss0})|\varepsilon(t)| =\cr
  &{}=
  (t-t_{\sss0})\,\hbox{dist}
  \bigl(p(t_{\sss0},\pqbar),\partial\D
  \bigr)-
  (t-t_{\sss0})|\varepsilon(t)|\,.\cr}
  \eqno(3.9)$$
If we apply this inequality first for
$t_{\sss0}=0$, reminding that
dist$(\bar p,\partial\D)\ge2\gamma$, we obtain
$$\psi(t)>\gamma t\quad\hbox{ for all $t>0$ small
  enough.}
  \eqno(3.10)$$
On the other hand, for all $t_{\sss0}\ge0$ where
$\psi$ is differentiable, we can write
$$\psi^\prime(t_{\sss0})\ge
  \hbox{dist}\bigl(p(t_{\sss0},\pqbar),
  \partial\D\bigr)\,.
  \eqno(3.11)$$
For almost all $t$ in any interval where $\psi$
is defined we have
$$\eqalign{\psi^\prime(t)&{}\ge
  \hbox{dist}\bigl(p(t,\pqbar),
  \partial\D\bigr)=\cr
  &{}=
  \min_{v\in\bar\C\atop|v|=1}
  \Bigl(\bar p\cdot v+
  \int_0^t-\nabla\V(q(s,\bar p,\bar q))\cdot
  v\,ds\Bigr)\ge\cr
  &{}\ge
  \min_{v\in\bar\C\atop|v|=1}\bar p\cdot v-
  \int_0^th_{\sss0}\bigl(
  \hbox{dist}(q(s,\bar p,\bar q),q_{\sss0}+
  \partial\D)\bigr)\,ds\ge\cr
  &{}\ge
  2\gamma-\int_0^t
  h_{\sss0}\bigl(\psi(s)+b_\gamma
  \bigr)\,ds\,.\cr}
  \eqno(3.12)$$

\noindent 
Integrating the inequality we get (recalling that
$\psi(0)\ge0$):
$$\psi(t)\ge2\gamma t-\int_0^t(t-s)\,
  h_{\sss0}\bigl(\psi(s)+b_\gamma\bigr)\,ds\,.
  \eqno(3.13)$$
For the function $\tilde\psi(t):=\gamma t$ we
have 
$$\eqalign{2\gamma t-\int_0^t(t-s)\,
  h_{\sss0}\bigl(\tilde
  \psi(s)+b_\gamma\bigr)\,ds
  ={}&\cr
  =2\gamma t-\int_0^t(t-s)\,
  h_{\sss0}\bigl(
  \gamma s+b_\gamma\bigr)\,ds
  \ge{}&\cr
  \ge2\gamma t-t\int_0^th_{\sss0}(
  \gamma s+b_\gamma)\,ds\ge{}&\cr
  \ge2\gamma t-t\int_0^{+\infty}\mskip-14mu
  h_{\sss0}(\gamma s+b_\gamma)\,ds>{}&
  2\gamma t-\gamma t=\gamma t=\tilde\psi(t)\,.
  \cr}
  \eqno(3.14)$$

\noindent
Since $\psi(t)>\tilde\psi(t)$ for small
enough~$t>0$, a standard argument in integral
inequalities (see e.g.~[LL], Theorem~5.5.1)
yields 
$$\psi(t)\ge\gamma t\quad\forall t\ge0\,.
  \eqno(3.15)$$
The relation dist$(p(t,\bar p,\bar q),
\partial\D))\ge\gamma$ comes from
$$\hbox{dist}\bigl(p(t,\bar p,\bar q),
  \partial\D\bigr)\ge
  2\gamma-\int_0^th_{\sss0}\bigl(\psi(s)+b_\gamma
  \bigr)\,ds\ge\gamma\,.
  \eqno(3.16)$$
\qed

\goodbreak
\bigskip

Proving that $\M$ is open is easy now.

\bigskip

{\bf Proposition 3.3 } \sl Suppose that
Hypotheses~2.1 and~3.1 are verified. Then,
for each $(\pqzero)\in\Rn\times
\hbox{{\rm dom}}\V$ whose asymptotic velocity
$\pinf(\pqzero)$ belongs to the interior of the
dual cone~$\D$ and each $q^\prime\in\hbox{{\rm
dom}}\V$, there exist $\gamma>0$,
$t_{\sss0}\in\R$ and a bounded neighbourhood $U$
of $(\pqzero)$ in $\Rn\times\dom\V$ such that,
for all $t\ge t_{\sss0}$ and $(\pqbar)\in U$ we
have \rm 
$$\matrix{p(t,\pqbar)\in\D^\circ,\hfill&
  \hbox{dist}\bigl(p(t,\pqbar),\partial\D
  \bigr)\ge\gamma,\hfill\cr
  \noalign{\smallskip}
  q(t,\pqbar)\in q^\prime+\D,\hfill&
  \hbox{dist}\bigl(q(t,\pqbar),q^\prime+
  \partial\D\bigr)\ge
  \gamma(t-t_{\sss0}).\hfill\cr}
  \eqno(3.17)$$
\sl In particular, $\pinf(\pqbar)\in\D^\circ$ for
all $(\pqbar)\in U$, so that $\M$ is open
in~$\Rn\times\Rn$. \rm

\goodbreak
\bigskip

{\bf Proof. } Let $4\gamma:=\hbox{dist}(\pinf(
\pqzero),\partial\D)>0$. The trajectory
$q(t,\pqbar)$ eventually enters all sets of the
form $q+\D^\circ$. Let $t_{\sss0}\in\R$ be such
that 
$$\hbox{dist}(p(t_{\sss0},\pqzero),
  \partial\D)\ge3\gamma\,, \quad
  q(t_{\sss0},\pqzero)\in
  (q_\gamma+\D^\circ)\cap
  (q^\prime+\D^\circ)\,.
  \eqno(3.18)$$
Let $U$ be a bounded neighbourhood of
$(\pqzero)$ such that, for all $(\pqbar)\in U$
we have
$$p(t_{\sss0},\pqbar)\in\D^\circ\,,\quad
  \hbox{dist}\bigl(p(t_{\sss0},\pqbar),
  \partial\D\bigr)\ge2\gamma\,,\quad
  q(t_{\sss0},\pqbar)\in(q_\gamma+\D^\circ)
  \cap(q^\prime+\D^\circ)\,.
  \eqno(3.19)$$
To conclude we only need to apply
Proposition~3.2. \qed

\goodbreak
\bigskip

{\bf Proposition 3.4 } \sl Suppose that the
Hypotheses~1.1 and~3.1 are
verified. Then the mapping
$(\pqbar)\mapsto\pinf(\pqbar)$ is continuous
from~$\M$ onto all of~$\D^\circ$. If, moreover,
the function $h_{\sss0}$ is integrable
on~$[0,+\infty[$, then the asymptotic map $\A$
defined in~(1.5) exists on~$\M$, it is
continuous and its range is all of $\D^\circ
\times\Rn$. \rm

\goodbreak
\bigskip

{\bf Proof. } See [GZ1], Section~5 and
Proposition~7.5 and [GZ2], Section~2. \qed

\goodbreak
\bigskip

We are now going to write the two sets
of assumptions on the potential~$\V$ that
in~[GZ1] and~[GZ2] were shown to imply the
smoothness of the asymptotic data in~$\M$.

We will denote by~$\Dif^k\V$ the $k$-th
differential of~$\V$, regarded as a multilinear
map from $(\Rn)^{k-1}$ into~$\Rn$, endowed
with the norm
$$\|\Dif^k\V(q)\|:=\sup\{
  |\Dif^k\V(q)(x^{(1)},\ldots,x^{(k-1)})|\;:\;
  x^{(j)}\in\Rn,\;|x^{(j)}|\le1\}.$$

\bigskip

{\bf Hypotheses 3.5} \sl The potential $\V$ is a
$C^{m+1}$, $m\ge2$, function. For all
$1\le i\le m$ there exist $q_i\in\Rn$,
$A_i\ge0$, $\lambda_i>0$ such
that  
$$q\in q_i+\D\quad\Rightarrow\quad
  \|\Dif^{i+1}\V(q)\;\|\le
  A_i\exp\Bigl(-\lambda_i\,
  \hbox{{\rm dist}}\bigl(
  q,q_i+\partial\D\bigr)
  \,\Bigr)\,\hbox{,}
  \eqno(3.21)$$
\rm

\goodbreak
\bigskip

{\bf Hypotheses 3.6} \sl The potential $\V$ is a
$C^{m+1}$, $m\ge2$, function. For all
$1\le i\le m$ there exist $q_i\in\Rn$, and a
weakly decreasing function
$h_i\colon[0,+\infty[\to\R$ such that  
\item{1)} $\V$ is convex on
  $q_{\sss1}+\D$; 
\item{2)} for all
  $q^\prime,\;q^{\prime
  \prime}\in q_{\sss1}+\D$ and all $z\in\Rn$
  we have
  $$q^{\prime\prime}\in q^\prime+\D\quad
    \Rightarrow\quad
    \Dif^2\V(q^{\prime\prime})z\cdot z\le
    \Dif^2\V(q^\prime)
    z\cdot z\hbox{;}$$
\item{3)} for all $i$,
  $\int_0^{+\infty}\mskip-7mu 
  x^i\, h_i(x)dx<+\infty$ and
  $$q\in q_i+\D\quad\Rightarrow\quad
    \|\Dif^{i+1}\V(q)\|\le h_i\Bigl(\,
    \hbox{{\rm dist}}\bigl(q,q_i+
    \partial\D\bigr)\,\Bigr)\hbox{;}$$

\rm

\goodbreak
\bigskip

Here is the precise statement of the
integrability.

\bigskip

{\bf Proposition 3.7} \sl Assume Hypotheses~2.1,
3.1 and either~3.5 or~3.6. Then the components
of the asymptotic velocity~$\pinf$ are $C^m$
first integrals, independent and in involution
on the set~$\M$ of the asymptotically regular
initial data. If, moreover, the functions $h_i$
of Hypotheses~3.1 and~3.6 verify $\int_0^{+\infty}
x^{i+1}\,h_i(x)\,dx<+\infty$ for all $0\le i\le
m+1$, then the asymptotic map~$\A$ exists
on~$\M$, it is a global canonical
$C^m$ diffeomorphism from~$\M$ onto
$\D^\circ\times\Rn$, and it brings the restricted
phase space system~(1.8) into the normal form
$\dot P=0$, $\dot Q=P$. \rm

\bigskip

Let a potential $\V$ verify all the assumptions
of the previous proposition, and let us modify
it smoothly {\it outside a set of the form}
$q+\D$. Then all that happens to the thesis of
the proposition is that the set~$\M$ has changed.
We will see in the next section how sufficiently
small perturbations of~$\V$ may not even
alter~$\M$.

      \vfill\eject

%%%%%%%%%%%%%%%%%
\centerline{{\bfuno 4. Finite-sum Potentials}}
\bigskip

In this Section we concentrate our attentions on
the potentials which can be written as finite
sums of one-dimentional functions (``finite-sum
potentials''):
$$\V(q):=\sum_{\alpha=1}^N
  f_\alpha(q\cdot v_\alpha)\,,
  \eqno(4.1)$$
where $v_1,\ldots,v_N$ are given nonzero vectors
in~$\Rn$ ($N\ge1$, no relation to $n$), and the
functions $f_1,\ldots,f_N$ are real functions of
one variable.

Sufficient conditions, under which
Proposition~3.7 applies to the potential~$\V$
with its associated Hamiltonian system, are as
follows.

\bigskip

{\bf Hypotheses 4.1 } \sl The $f_\alpha$ are
$C^{m+1}$ ($m\ge2$) functions, whose domains 
$\hbox{{\rm dom}}\,f_\alpha$ are each
either~$\R$ or the interval $]0,+\infty[$,  and
$$\eqalignno{&\sup f_\alpha=+\infty\,,\qquad
  \inf f_\alpha=0\,,&(4.2)\cr
  &f_\alpha^\prime(x)<0\quad\forall x\in
  \dom\,f_\alpha\,,
  &(4.3)\cr
  &f_\alpha^{(k)}(x)\cases{
     >0&if $k$ is even,\cr
     <0&if $k$ is odd,\cr}
                        \quad\forall x\ge a\,,
  &(4.4)\cr
  &f_\alpha^{(m+1)} \hbox{ is monotone on
           $[a,+\infty[$,}
  &(4.5)\cr}$$
where $a>0$ is a constant. Finally, the vectors
$v_1,\ldots,v_N$ are nonzero and the cone
generated by them is proper. \rm

\goodbreak
\bigskip

The domain of $\V$ is  either $\Rn$ or the set
$$\{q\in\Rn\;:\;q\cdot v_\alpha\in \dom
  f_{\alpha}\quad
  \forall\alpha=1,\ldots,N\}\,.
  \eqno(4.6)$$
As we already observed in the Introduction, the
convex cone $\C$ generated by the forces
$-\nabla\V(q)$ is contained in the one generated
by the vectors $v_1,\ldots,v_N$, and for our
potentials the two have the same closure.

\goodbreak
\bigskip

{\bf Proposition 4.2 } \sl If Hypotheses~4.1
hold, the closure $\bar\C$ of $\C$
coincides with the convex cone generated by \rm
$v_{\sss1},\ldots,v_{\sss N}$:
$$\bar\C=\biggl\{ -\sum_{\alpha=1}^N
  \lambda_\alpha v_\alpha\;:\;
  \lambda_\alpha\ge0
  \biggr\}.
  \eqno(4.7)$$

\goodbreak
\bigskip

{\bf Proof. } See [GZ1], Lemma 10.2.
\qed

\bigskip

The dual cone $\D$ of $\bar\C$ is then
$$\D:=\{w\in\Rn\;:\;w\cdot v\ge0\quad\forall
                    v\in\C\}=
      \{w\in\Rn\;:\;w\cdot v_\alpha\ge0
        \quad\forall\alpha=1,\ldots,N\}\,.
  \eqno(4.8)$$
The set in formula (4.6) is simply
the interior of~$\D$.

\goodbreak
\bigskip

Hypotheses 2.1 are certainly verified for our
potentials. Also Hypo\-thesis~3.1 holds with
$q_{\sss0}$ any point in~$\D^\circ$ whose
distance from~$\partial\D$ is~$\ge a$, and
$h_{\sss0}$ given by
$$h_{\sss0}(x):=\sum_{\alpha=1}^N
  |v_\alpha|\;\bigl|
  f_\alpha(x\min_{\beta}|v_\beta|+a)\bigr|\,.
  \eqno(4.9)$$
Hypotheses~3.6 hold as well and the verification
can be found  in~[GZ1], Section~10. We can sum up
with the following statement.

\bigskip

{\bf Proposition 4.3 } \sl If Hypotheses~4.1
hold, then the first part of Proposition~3.7
applies to the potential~(4.1). If moreover
all the functions $f_\alpha$ have finite
integrals on $[a,+\infty[$, then also the second
part applies. \rm

\goodbreak
\bigskip

To study the effects of small perturbations of
the potential~$\V$, we start with some technical
Lemmas.

\bigskip

{\bf Lemma 4.4 } \sl If the nonzero vectors
$v_{\sss1},\ldots,v_{\sss N}$ generate a proper
cone~$C$ in~$\Rn$, (i.e., the dual $D$ of~$C$ has
nonempty interior), then there exists a vector
$\bar v\in D^\circ$ of the form
$$\bar v=\sum_{\alpha=1}^N\mu_\alpha v_\alpha
  \quad
  \hbox{ with $\mu_\alpha>0$ for all $\alpha$.}
  \eqno(4.10)$$
\rm

\goodbreak
\bigskip

{\bf Proof. } Let us first prove that
$C\cap D^\circ$ is nonempty. In fact, if that
were not the case, we could separate $C$
from~$D$, i.e., there would exist $\bar w\ne0$
such that
$$\eqalign{&\bar w\cdot v_\alpha\ge0
  \qquad\forall\alpha=1,\ldots,N\,,\cr
  &\bar w\cdot v\le0\qquad
  \forall v\in D\,.\cr}
  \eqno(4.11)$$
The first relation is equivalent to $\bar w\in
D$, and this, together with the second one,
would yield $|\bar w|^2=\bar w\cdot\bar w\le0$,
that contradicts~$\bar w\ne0$.

\noindent
Let us now pick
$\bar v^\prime\in C\cap D^\circ$. In particular,
$\bar v^\prime$ is of the form
$$\bar v^\prime=\sum_{\alpha=1}^N
  \lambda_\alpha v_\alpha\,,\qquad
  \lambda_\alpha\ge0\,,
  \eqno(4.12)$$
and $\bar v^\prime$ belongs to the {\it open}
set~$D^\circ$. Choose $\tau>0$ small enough for
the vector
$$\bar v:=\bar v^\prime+\tau\sum_{\alpha=1}^N
  v_\alpha=\sum_{\alpha=1}^N(\lambda_\alpha+\tau)
  v_\alpha
  \eqno(4.13)$$
to belong to $D^\circ$. This is what we were
looking for. \qed

\goodbreak
\bigskip

{\bf Lemma 4.5 } \sl Let $v_{\sss1},\ldots,
v_{\sss N}$, $C$, $D$, $\bar v$ be as in
Lemma~4.4. Then, for all $\bar q\in\Rn$ and
$c>0$, each of the functions $q\mapsto q\cdot
v_\alpha$ is bounded on the set \rm
$$\{q\in\Rn\;:\; q\in\bar q+D,\quad
  q\cdot\bar v\le c\}\,.
  \eqno(4.14)$$

\goodbreak
\bigskip

{\bf Proof. } From $q\in\bar q+D$ we see that
$q\cdot v_\alpha$ is bounded from below: $\bar
q\cdot v_\alpha\le q\cdot v_\alpha$. The
condition $q\cdot\bar v\le c$ is
$$\sum_{\alpha=1}^N\mu_\alpha q\cdot v_\alpha
  \le c\,.
  \eqno(4.15)$$
Each term $\mu_\alpha q\cdot v_\alpha$ is
bounded from below, and their sum is bounded from
above. Hence all of them are bounded from above,
too. Remind finally that $\mu_\alpha>0$ for
all~$\alpha$. \qed

\goodbreak
\bigskip

{\bf Lemma 4.6 } \sl In the Hypotheses~4.1,
let the vector $\bar v$ be supplied by
Lemma~4.4. Then, for all $\bar q\in\dom\V$
and $c>0$ we have \rm
$$\inf\{-\nabla\V(q)\cdot\bar v\;:\;
  q\in\bar q+\D\,,\;
  q\cdot\bar v\le c\}>0\,.
  \eqno(4.16)$$

\goodbreak
\bigskip

{\bf Proof. } Let $b$  be an upper bound for
$q\cdot v_{\alpha_0}$ on $\{q\;:\;q\in\bar
q+\D,\; q\cdot\bar v\le c\}$. Then, for $q$ in
that set,
$$\eqalign{-\nabla\V(q)\cdot\bar v={}&
  -\sum_{\alpha=1}^Nc_\alpha
  f^\prime_\alpha(q\cdot
  v_\alpha)v_\alpha\cdot\bar v\ge
  -c_{\alpha_0}f^\prime_{\alpha_0}
  (q\cdot v_{\alpha_0})
  v_{\alpha_0}\cdot\bar v\ge\cr
  \ge{}&c_{\alpha_0}v_{\alpha_0}\cdot\bar v\,
  \min\{-f^\prime_{\alpha_0}(x)\;:\;
  \bar q\cdot v_{\alpha_0}\le x\le b\}>0\,.\cr}
  \eqno(4.17)$$
\qed

\goodbreak
\bigskip

{\bf Lemma 4.7 } \sl Assume Hypotheses~4.1. 
Then for all $E>0$ there exists $q_{\sss
E}\in\dom\V$ such that \rm
$$\V(q)\le E
  \quad\Rightarrow\quad
  q\in q_{\sss E}+\D\,.
  \eqno(4.18)$$

\goodbreak
\bigskip

{\bf Proof. } The function $x\mapsto 
f_\alpha(x)$ is strictly decreasing and onto
$]0,+\infty[$. Define $q_{\sss E}:=\theta\bar v$,
where
$$\theta:=\min_\alpha
  {f_\alpha^{-1}(E)\over
  \bar v\cdot v_\alpha}.
  \eqno(4.19)$$
We have $q_{\sss E}\cdot v_\alpha\le
f_\alpha^{-1}(E)$ for all~$\alpha$. If $q\in
\dom\V$ is such that
$q\notin q_{\sss E}+\D$, then there 
exists~$\alpha$ for
which $q\cdot v_\alpha<
q_{\sss E}\cdot v_\alpha$ and
hence
$$\V(q)\ge f_\alpha(q\cdot v_\alpha)\ge
  f_\alpha(q_{\sss E}\cdot v_\alpha)\ge E\,.
  \eqno(4.20)$$
\qed

\goodbreak
\bigskip

Here is what we can say at this point about the
asymptotic behaviour of the trajectory of a
slightly perturbed system. 

\bigskip

{\bf Proposition 4.8 } \sl Suppose that
Hypotheses~4.1 hold for the potential~(4.1). Let
$K$ be a compact set contained in the domain
of~$\V$. Then there exists an $\varepsilon>0$
with the following property. Let $V$ be a
$C^{m+1}$ real function defined in~$\dom\V$,
vanishing outside~$K$, and such that
$$\sup_K|\nabla V|\le\varepsilon\,.
  \eqno(4.21)$$
Then each motion of the Hamiltonian system
associated with the perturbed potential $\V+V$
coincides on an interval $[t_{\sss0},+\infty[$
with a motion of the original system. \rm

\goodbreak
\bigskip

{\bf Proof. } Let $\bar q\in\dom\V$ be such that
$K\subset\bar q+\D$. Let $\bar v$ be given by
Lemma~4.4. We claim that the thesis holds for
any $\varepsilon$ such that
$$0<\varepsilon<
  \inf\{-\nabla\V(q)\cdot
  \bar v\;:\;
  q\in K\}\,,
  \eqno(4.22)$$
(the infimum is $>0$ because of Lemma~4.6).
In fact, let $(\tilde p(t),\tilde q(t))$ be a
trajectory of the perturbed system. We will show
that $\tilde q(t)\cdot\bar v\to+\infty$ as
$t\to+\infty$, so that $\tilde q(t)$ eventually
quits the compact set~$K$ forever.

\noindent
There certainly exists  $q_{\sss E}\in\Rn$ such
that $\tilde q(t)\in q_{\sss E}+\D$ for all
$t\in\R$ because the conservation of
energy and the thesis of Lemma~4.7 still hold
for $\V+V$. 

\noindent
The function $t\mapsto\tilde q(t)\cdot\bar v$ is
convex: $\ddot{\tilde q}(t)\cdot\bar v=
-\nabla(\V+V)(\tilde q(t))\cdot\bar v>0$ for
all~$t\in\R$ because of~(4.21) and~(4.22).
If it did not tend to~$+\infty$, as
claimed, then it would be decreasing, and
$\tilde q(t)$ would belong to $\{q\in\Rn\;:\;
q\in q_{\sss E}+\D,\; q\cdot\bar v\le \tilde
q(0)\cdot\bar v\}$ for
all~$t\ge0$.

\noindent
Let 
$$\varepsilon^\prime:=\inf
  \bigl\{-\nabla\V(q)\cdot \bar v\;:\;
  q\in q_{\sss E}+\D\,,\;
  q\cdot\bar v\le 
  \tilde q(0)\cdot\bar q\bigr\}>0
  \eqno(4.23)$$
(see Lemma~4.6). We can write
$$\Bigl(\; q\in q_{\sss E}+\D\hbox{ and }
  q\cdot\bar v\le \tilde q(0)\cdot\bar q\;\Bigr)
  \;\;\Rightarrow\;\;
  -\nabla(\V+V)(q)\cdot\bar v\ge
  \min\{\varepsilon-\sup|\nabla V|\,,\,
  \varepsilon^\prime\}>0\,.
  \eqno(4.24)$$
But then $t\mapsto\tilde
q(t)\cdot\bar v$ cannot be decreasing, because
its second derivative is greater than a
positive constant for all~$t\ge0$.
\qed

\goodbreak
\bigskip

Suppose that for a finite-sum potential verifying
Hypotheses~4.1 the set~$\M$ of the
asymptotically regular initial data coincides
with the whole phase space. Then the associated
Hamiltonian system is globally integrable {\it
together with all slightly perturbed systems} as
specified by the previous proposition.

\goodbreak

The paper~[GZ1] provides a class of globally
integrable finite-sum potential systems. At this
point it is easily described:

\bigskip

{\bf Proposition 4.9 } \sl Assume the
Hypotheses~4.1 and that the vectors
$v_1,\ldots,v_N$ verify
$$v_\alpha\cdot v_\beta\ge0\qquad
  \forall\alpha,\beta=1,\ldots,N\,.
  \eqno(4.25)$$
Then for the associated Hamiltonian System all
initial data are asymptotically regular. \rm

\goodbreak
\bigskip

The proof of Proposition~4.8 contains
also the proof of the following one.

\bigskip

{\bf Proposition 4.10 } \sl Assume the
Hypotheses~4.1. Then, for all initial data
$(\pqbar)$, 
$$\pinf(\pqbar)\in\D\,.
  \eqno(4.26)$$
Moreover, let the vector $\bar v$
be given by Lemma~4.4. Then 
$$\pinf(\pqbar)\cdot\bar v>0\,.
  \eqno(4.27)$$
In particular, the asymptotic velocity never
vanishes. \rm

\goodbreak
\bigskip

{\bf Proof. } Formula~(4.26) comes from the
fact that $q(t,\pqbar)$ belongs to a set of
the form $q_{\sss E}+\D$ for all~$t$. 
Formula~(4.27) comes along with the argument
used in the proof of Proposition~4.8, assuming
$V\equiv0$. \qed

      \vfill\eject

%%%%%%%%%%%%%%%%%
\centerline{{\bfuno 5. A Counterexample to
             Global Asymptotic Regularity}}  
\bigskip

Let us start with a finite-sum potential
in~$\R^2$ of the form
$$\V(x,y):=2e^{-x+y}+f(y)\,,\qquad
  (x,y)\in\R^2,
  \eqno(5.1)$$
where $f\colon\R\to\R$ is a smooth function to be
determined. We impose that the following
trajectory
$$\cases{x(t):=4t-2\ln t\,,&\cr
  y(t):=4t-4\ln t&\cr}
  \eqno(5.2)$$
be a solution, for large~$t$, of the 
Hamiltonian system associated to~$\V$. We are
going to see that this is indeed possible with
$\V$ satisfying Hypotheses~4.1. The cone of the
forces will turn out to be
$$\C=\{(x,y)\in\R^2\;:\;y>0,\;x+y>0\}
  \eqno(5.3)$$
which is generated by the vectors $(1,-1)$ and
$(0,1)$, and is {\it wider than~$\pi/2$}. The
dual cone will be
$$\D=\{(x,y)\in\R^2\;:\;y\ge0,\;x-y\ge0\}\,.
  \eqno(5.4)$$
The asymptotic velocity for the
trajectory~(5.2) is the vector $(4,4)\in\R^2$,
which is {\it on the boundary of~$\D$}.

The acceleration of the trajectory is
$$\ddot x(t)={2\over t^2}\,,\qquad
  \ddot y(t)={4\over t^2}\,.
  \eqno(5.5)$$
The force $-\nabla\V$ has components
$$-{\partial\V\over\partial x}(x,y)
  =2e^{-x+y},\qquad
  -{\partial\V\over\partial y}(x,y)=
  -2e^{-x+y}-f^\prime(y)\,.
  \eqno(5.6)$$
We must impose the equality $(\ddot x,\ddot y)
=-\nabla\V$. For the $x$~component this is
already true:
$${2\over t^2}=\ddot x(t)=-{\partial\V\over
  \partial x}(x(t),y(t))=2e^{-x(t)+y(t)}=
  2e^{-2\ln t}.
  \eqno(5.7)$$
For the $y$ component we get the following
condition on~$f$:
$${4\over t^2}=\ddot y(t)=-{\partial\V\over
  \partial y}(x(t),y(t))=
  -{2\over t^2}+f^\prime(y(t))\,,
  \eqno(5.8)$$
that is,
$$f^\prime(4t-4\ln t)=-{6\over t^2}\,.
  \eqno(5.9)$$
Upon multiplication by $4-4/t$ and integration we
get 
$$f(4t-4\ln t)={24\over t}-{12\over t^2}+c\,.
  \eqno(5.10)$$
The function $t\mapsto4t-4\ln t$ is a $C^\infty$
diffeomorphism between the intervals, say,
$[2,+\infty[$ and~$[8-4\ln2,+\infty[$. So
(selecting the constant $c=0$) there exists a
$C^\infty$ function $f\colon\R\to\R$ such that 
$$\eqalignno{&\sup f=+\infty\,,\qquad
  \inf f=0\,,
  &(5.11)\cr
\noalign{\smallskip}
  &f^\prime(y)<0\quad\forall y\in\R\,,
  &(5.12)\cr
  &f(4t-4\ln t)={24\over t}-{12\over t^2}
  \qquad\forall t\ge2\,.
  &(5.13)\cr}$$
The function $f$ is {\it not integrable
at~$+\infty$}:
$$\int_{8-4\ln2}^{4t-4\ln t}\mskip-14mu
  f(y)\,dy=
  \int_2^t f(4s-4\ln s)\Bigl(4-{4\over s}\Bigr)
  ds=
  96\int_2^t\Bigl({1\over s}-{3\over2s^2}+
  {1\over2s^3}\Bigr)ds\,,
  \eqno(5.14)$$
which diverges as $t\to+\infty$. Of course,
$f$~is not a multiple of an exponential
function, not even asymptotically. The cones
$\C$ and~$\D$ are easily verified to be as
announced.

We are only left to prove that the derivatives
of~$f$ have alternate signs:
$$f^{(k)}(y)\cases{>0&if $k$ is even,\cr
  <0&if $k$ is odd\cr}
  \eqno(5.15)$$
for all $k$ and all large $y$ (possibly
depending on~$k$). The inequalities already
hold, globally, for~$k=0,\,1$.
Differentiating~(5.9) we get
$$f^{(k)}(4t-4\ln t)=\varphi_k(t)\,,
  \eqno(5.16)$$
where the function $\varphi_k$ is defined
recursively as
$$\varphi_1(t):=-{6\over t^2}\,,\qquad
  \varphi_{k+1}(t):=\varphi_k^\prime(t)
  {t\over4t-4}\,.
  \eqno(5.17)$$
It is easy to see that $\varphi_k$ is a rational
function and that the degree of the denominator
exceedes (by~$k+1$) the degree of the numerator.
Then $\varphi_k$ is monotone and
infinitesimal at~$+\infty$, so that
$\varphi_k^\prime(t)$, and hence
$\varphi_{k+1}(t)$ too, has the opposite sign
of~$\varphi_k(t)$ for all large~$t$.

We can conlude with the following statement: \sl
For the Hamiltonian system associated with this
cone potential~$\V$, the two components of the
asymptotic velocity are $C^\infty$ integrals of
motion, independent and in involution on the
nonempty, open, invariant set~$\M$ of the
asymptotically regular initial data. However,
$\M$~does not coincide with the whole phase
space. \rm

      \vfill\eject

%%%%%%%%%%%%%%%%%
\centerline{{\bfuno 6. The Case of All Equal
                     Functions}} 
\bigskip

In this Section we make the
following assumptions on the
functions~$f_\alpha$ of Section~4.

\bigskip

{\bf Hypothesis 6.1 } \sl All the functions
$f_\alpha$ are multiples of a single smooth
function~$f$:
$$f_\alpha=c_\alpha f\,,\quad c_\alpha>0\,,
  \eqno(6.1)$$
defined on either $\R$ or on $]0,+\infty[$,
and such that $f^\prime(x)<0$ for all~$x$ and
$$x\mapsto x|f^\prime(x)| 
  \hbox{ is weakly  decreasing on }[a,+\infty[
  \eqno(6.2)$$
for some $a>0$.
\rm

\goodbreak
\bigskip

{\bf Proposition 6.2 } \sl Suppose that
Hypotheses~4.1 and~6.1 hold. Then, for the
Hamiltonian system associated to the
potential~$\V$, all initial data $(\pqbar)
\in\dom\V$ are asymptotically regular, i.e.,
the asymptotic velocity always belongs to the
interior of the dual cone~$\D$:  
$$\pinf(\pqbar)\cdot v_\alpha>0
  \qquad\forall(\pqbar)\in\dom\V\,,
  \quad\forall\alpha=1,\ldots,N\,.
  \eqno(6.3)$$
\rm

\goodbreak
\bigskip

The condition that $x\mapsto x|f^\prime(x)|$ be
monotone is not very restrictive. With only
Hypo\-theses~4.1, the derivative $|f^\prime|$ is
monotone and integrable on~$[a,+\infty[$, and
this already implies that $x|f^\prime(x)|\to0$ as
$x\to+\infty$ (this is elementary; see~[GZ1],
Lemma~10.3). Examples of functions~$f$ that
verify our requirements are
$$\eqalignno{&f(x):=e^{-x},\qquad
  x\in\R\,;&(6.4)\cr
  &f(x):={1\over x^r}\,,\qquad
  x>0\,,\quad r>0\,;&(6.5)\cr
  &f(x):={1\over(\ln(1+x))^r}\,,\qquad
  x>0\,,\quad r>0\,.
  &(6.6)\cr}$$

\goodbreak
\bigskip

{\bf Lemma 6.3 } \sl Suppose that $f$ verifies
the Hypo\-thesis~6.1. Let $\varphi\colon\R\to
\dom f\subset\R$
be a function such that
$$\inf\varphi=L>\inf\,\dom f\,,\qquad
  \lim_{t\to+\infty}{\varphi(t)\over t}=0\,.
  \eqno(6.7)$$
Then, for any $\gamma>0$, \rm
$$\lim_{t\to+\infty}{f^\prime(\gamma t)\over
  f^\prime(\varphi(t))}=0\,.
  \eqno(6.8)$$

\bigskip

{\bf Proof. } For all large~$t$ we have $\gamma
t\ge\varphi(t)$ and $\gamma t\ge a$. Then, using
the monotonicity of $x\mapsto x|f^\prime(x)|$:
$$0<|f^\prime(\gamma t)|=
  {1\over\gamma t}\gamma t|f^\prime(\gamma t)|
  \le\theta(t)|f^\prime(\varphi(t))|\,,
  \eqno(6.9)$$
where $\theta(t)$ is defined as
$$\theta(t):=\cases{
  \displaystyle{{\varphi(t)\over\gamma t}}&
  if $\varphi(t)\ge a$,\cr
  \noalign{\smallskip}
  \displaystyle{{a|f^\prime(a)|\over
  \gamma t\min\{|f^\prime(x)|\;:\;
  L\le x\le a\} }}&
  if $\varphi(t)\le a$.\cr}
  \eqno(6.10)$$
It is clear that $\theta(t)\to0$ as
$t\to+\infty$. \qed

\goodbreak
\bigskip

{\bf Proof of Proposition 6.2 } We suppress
the initial conditions $(\pqbar)$ from the
notation, because we are only interested in
single trajectories. So, let $(p(t),q(t))$ be a
motion of the system, with asymptotic
velocity~$\pinf$. We already know (Lemma~4.7)
that $q(t)\in q_{\sss E}+\D$ for all~$t$, with
$q_{\sss E}\in\dom\V$, and that $0\ne\pinf\in\D$
(Proposition~4.10).  What we are left to prove is
that $\pinf\notin\partial\D$. 

\noindent
Suppose the contrary.
Let $I_0$ be the subset of $\{1,\ldots,N\}$ where
$\pinf\cdot v_\alpha=0$ and $I_1$ be the
complement, i.e., where $\pinf\cdot v_\alpha>0$.
The set $I_0$ is nonempty because
$\pinf\in\partial\D$. The complement $I_1$~is
nonempty because $0\ne\pinf\in\D$. 

\noindent
With Lemma~4.4 applied to $\{v_\alpha\;:\;
\alpha\in I_0\}$, we can find a vector $\bar v$
such that
$$\eqalign{&\bar v=\sum_{\alpha\in I_0}
  \rho_\alpha v_\alpha\qquad 
  \rho_\alpha>0\,,\cr
  &\bar v\cdot v_\alpha>0\qquad
  \forall \alpha\in I_0\,.\cr}
  \eqno(6.11)$$
In particular, $\bar v$ is orthogonal to~$\pinf$.

\noindent
Define the function
$$g(t):=q(t)\cdot\bar v\,.
  \eqno(6.12)$$
We have $g^\prime(t)=p(t)\cdot\bar v\to\pinf
\cdot\bar v=0$. We will reach a contradiction by
showing that
$g^{\prime\prime}(t)\ge\hbox{constant}>0$
for large~$t$.

\noindent
Let us split $\V$ into the sum of two potentials
$$\V_0(q):=\sum_{\alpha\in I_0}
  c_\alpha f(q\cdot v_\alpha)\,,\qquad
  \V_1(q):=\sum_{\alpha\in I_1}
  c_\alpha f(q\cdot v_\alpha)\,.
  \eqno(6.13)$$
The gradient
of~$\V_0$ is orthogonal to~$\pinf$, whilst
$\pinf$ is in the interior of the dual of the
cone spanned by~$-\nabla\V_1$.

\noindent
Let
$$2\gamma:=\min_{\alpha\in I_1}\pinf\cdot
  v_\alpha>0\,.
  \eqno(6.14)$$
We get $q(t)\cdot v_\alpha\ge\gamma t$ for all
large~$t$ and all $\alpha\in I_1$, so that
$$\eqalign{|\nabla\V_1(q(t))\cdot\bar v|
  \le{}&
  \sum_{\alpha\in I_1}c_\alpha|\bar v\cdot
  v_\alpha|
  \bigl|f^\prime(q(t)\cdot v_\alpha)
  \bigr|\le\cr
  \le{}&\sum_{\alpha\in I_1}
  c_\alpha|\bar v\cdot
  v_\alpha|\;
  \bigl|f^\prime(\gamma t)\bigr|:=
  c|f^\prime(\gamma t)|\cr}
  \eqno(6.15)$$
for all large $t$, because $|f^\prime|$ is
decreasing on~$[a,+\infty[$.

\goodbreak

\noindent
Let us consider the second derivative of~$g$:
$$\eqalign{g^{\prime\prime}(t)={}&
  -\nabla\V(q(t))\cdot\bar v=
  -\nabla\V_0(q(t))\cdot\bar v
  -\nabla\V_1(q(t))\cdot\bar v=\cr
  ={}&
  -\sum_{\alpha\in I_0}c_\alpha
  f^\prime(q(t)\cdot v_\alpha)\,
  v_\alpha\cdot\bar v
  -\nabla\V_1(q(t))\cdot\bar v\ge\cr
  \ge{}&
  -\sum_{\alpha\in I_0}c_\alpha
  f^\prime(q(t)\cdot v_\alpha)\,
  v_\alpha\cdot\bar v-
  c\,|f^\prime(\gamma t)|\cr}
  \eqno(6.16)$$
for all large $t$.

\goodbreak

\noindent
For any $\alpha\in I_0$, the
function $\varphi(t):=q(t)\cdot v_\alpha$ is
bounded from below by $q_{\sss E}\cdot v_\alpha$
and 
$$\lim_{t\to+\infty}{\varphi(t)\over t}=
  \lim_{t\to+\infty}\varphi^\prime(t)=
  \pinf\cdot v_\alpha=0
  \eqno(6.17)$$
with L'H\^opital's rule. We can apply
Lemma~6.3 and get that 
$$|f^\prime(\gamma t)|=
  o(-f^\prime(q(t)\cdot v_\alpha))
  \quad\hbox{as~$t\to+\infty$,}
  \eqno(6.18)$$
and, since  $-f^\prime>0$, 
$$g^{\prime\prime}(t)>0
  \quad\hbox{for all large $t$.}
  \eqno(6.19)$$

\noindent
But we know that $g^\prime(t)\to0$ as
$t\to+\infty$, so that $g$~turns out to be
decreasing for large~$t$:
$$q(t)\cdot\bar v\le q(t_0)\cdot\bar v\quad
  \hbox{for all $t\ge t_0$.}
  \eqno(6.20)$$
Next, Lemma~4.6, applied
to~$\V_0$ and~$\bar v$, together with
formula~(6.16), yields
$$-\nabla\V_0(q(t))\cdot\bar v\ge
  \inf\{-\nabla\V_0(q)\cdot\bar v\;:\;
  (q-q_{\sss E})\cdot v_\alpha\ge0\;
  \forall\alpha\in I_0,
  \;q\cdot\bar v\le 
  q(t_0)\cdot\bar v\}>0
  \eqno(6.21)$$
for all $t\ge t_0$, so that
$$g^{\prime\prime}(t)\ge\hbox{constant}>0\,,
  \eqno(6.22)$$
for all large $t$, which is the desired
contradiction. \qed

\bigskip

{\bf Note. } After reading Theorem~1 of~[Hu], we
realized that condition~(6.2) can be dropped.
In~fact, although equation~(6.18) may no longer
be true, it is still possible to write
$$\int_t^{+\infty}\mskip-14mu
  |f^\prime(\gamma s)|\,ds=
  o\biggl(
  \int_t^{+\infty}\mskip-14mu
  \bigl|f^\prime\bigl(q(s)\cdot v_\alpha\bigr)
  \bigr|\,ds\biggr)
  \quad\hbox{as $t\to+\infty$,}
  \eqno(6.23)$$
and apply it to the integrated version
of~(6.16). However, we have chosen to retain our
original proof, because it seems to use more
``mechanical'' quantities, and it applies to all
relevant examples we know of.

   \vfill\eject

%%%%%%%%%%%%%%%%%
\centerline{{\bfuno 7. The Case of all
                   Integrable Functions}} 
\bigskip

In this Section the assumptions on~$f_\alpha$,
in addition to Hypotheses~4.1, are of fast decay
type. Namely, they are integrable at~$+\infty$.

\bigskip

{\bf Hypothesis 7.1 } \sl For all
$\alpha=1,\ldots,N$, the function
$f_\alpha$ is such that \rm 
$$\int_a^{+\infty}\mskip-14mu
  f_\alpha(x)\,dx<+\infty\,.
  \eqno(7.1)$$

\goodbreak
\bigskip

{\bf Proposition 7.2 } \sl Suppose that
Hypotheses~4.1 and~7.1 hold. Then, for the
Hamiltonian system associated to the
potential~$\V$, all initial data $(\pqbar)
\in\dom\V$ are asymptotically regular, i.e.,
the asymptotic velocity always belongs to the
interior of the dual cone~$\D$:  
$$\pinf(\pqbar)\cdot v_\alpha>0
  \qquad\forall(\pqbar)\in\dom\V\,,
  \quad\forall\alpha=1,\ldots,N\,.
  \eqno(7.2)$$
\rm

\goodbreak
\bigskip

{\bf Lemma 7.3 } \sl If the Hypotheses 1.1
and~3.1 hold, then the function
$h_{\sss0}$ of formula~(4.9) verifies \rm
$$\int_0^{+\infty}
  x\,h_{\sss0}(x)\,dx<+\infty\,.
  \eqno(7.3)$$

\goodbreak
\bigskip

The statement of the following lemma is a
little awkward, because it is going to be
applied not to the original potential~$\V$.

\bigskip

{\bf Lemma 7.4 } \sl Let  $\tilde q\colon\R\to
\Rn$ be a $C^1$ function such that
$$\lim_{t\to+\infty}{d\tilde q\over dt}(t)
  =\tilde p_{\sss\infty}\in\D^\circ\,.
  \eqno(7.4)$$
Then the function $(r,s)\mapsto|\nabla\V(\tilde q
(s))|$ is integrable on
$\{(r,s)\in\R^2\;:\;0\le r\le s<+\infty\}$.
In particular, the following integrals converge
absolutely: \rm
$$\int_t^{+\infty}\mskip-8mu dr
  \int_r^{+\infty}\mskip-14mu
  \nabla\V(\tilde q(s))\,ds=
  \int_t^{+\infty}\mskip-14mu
  (s-t)\nabla\V(\tilde q(s))\,ds\,.
  \eqno(7.5)$$

\goodbreak
\bigskip

{\bf Proof. } If $2\gamma:=\hbox{dist}(\tilde
p_{\sss\infty},\partial\D)$, then the trajectory
satisfies 
$$\hbox{dist}\bigl(\tilde q(t),q(t_0)+\partial\D
  \bigr)\ge\gamma t\quad
  \hbox{for all $t\ge t_0$}
  \eqno(7.6)$$
for some $t_0$. We can conclude using Lemma~7.3.
\qed

\goodbreak
\bigskip

{\bf Proof of Proposition 7.2 } 
We suppress the initial conditions $(\pqbar)$
from the notation, because we are only
interested in single trajectories. So, let
$(p(t),q(t))$ be a motion of the system, with
asymptotic velocity~$\pinf$. We already know
(Lemma~4.7) that $q(t)\in q_{\sss E}+\D$ for
all~$t$, with $q_{\sss E}\in\dom\V$, and that
$0\ne\pinf\in\D$ (Proposition~4.10).  What we are
left to prove is that $\pinf\notin\partial\D$. 

\noindent
Suppose the contrary.
Let $I_0$ be the subset of $\{1,\ldots,N\}$ where
$\pinf\cdot v_\alpha=0$ and $I_1$ be the
complement, i.e., where $\pinf\cdot v_\alpha>0$.
The set $I_0$ is nonempty because
$\pinf\in\partial\D$. The complement $I_1$~is
nonempty because $0\ne\pinf\in\D$. 

\noindent
With Lemma~4.4 applied to $\{v_\alpha\;:\;
\alpha\in I_0\}$, we can find a vector $\bar v$
such that
$$\eqalign{&\bar v=\sum_{\alpha\in I_0}
  \rho_\alpha v_\alpha\qquad 
  \rho_\alpha>0\,,\cr
  &\bar v\cdot v_\alpha>0\qquad
  \forall \alpha\in I_0\,.\cr}
  \eqno(7.7)$$
In particular, $\bar v$ is orthogonal to~$\pinf$.

\noindent
Define the function
$$g(t):=q(t)\cdot\bar v\,.
  \eqno(7.8)$$
We have $g^\prime(t)=p(t)\cdot\bar v\to\pinf
\cdot\bar v=0$. We will reach a contradiction by
showing that
$g^{\prime\prime}(t)\ge\hbox{constant}>0$
for large~$t$.

\noindent
Let us split $\V$ into the sum of two potentials
$$\V_0(q):=\sum_{\alpha\in I_0}
  f_\alpha(q\cdot v_\alpha)\,,\qquad
  \V_1(q):=\sum_{\alpha\in I_1}
  f_\alpha(q\cdot v_\alpha)\,.
  \eqno(6.13)$$
The gradient
of~$\V_0$ is orthogonal to~$\pinf$, whilst
$\pinf$ is in the interior of the dual of the
cone spanned by~$-\nabla\V_1$.

\noindent
So far the proof was just the same as for
Proposition~6.2. Let us write $g(t)$ this way:
$$\eqalign{g(t)={}&
  \Bigl(q(t)+\int_t^{+\infty}\mskip-8mu
  dr\int_r^{+\infty}\mskip-14mu
  \nabla\V_1(q(s))\,ds\Bigr)\cdot\bar v-
  \Bigl(\int_t^{+\infty}\mskip-8mu
  dr\int_r^{+\infty}\mskip-14mu
  \nabla\V_1(q(s))\,ds\Bigr)\cdot\bar v:=\cr
  :={}&g_{\sss0}(t)-g_{\sss1}(t)\,.\cr}
  \eqno(6.14)$$
The integral converges because of Lemma~7.4
applied to~$\V_1$. It is obvious that
$g_1(t)\to0$ as $t\to+\infty$, and in particular
$g_1$~is bounded as $t\to+\infty$. 

\noindent
We are going to show that $g_0$ is bounded too.
In fact
$$\eqalign{g_0^\prime(t)={}&
  \Bigl(p(t)-\int_t^{+\infty}\mskip-14mu
  \nabla\V_1(q(s))\,ds\Bigr)\cdot\bar v
  \;\to\;\pinf\cdot\bar v=0\quad
  \hbox{ as $t\to+\infty$,}\cr
  g_0^{\prime\prime}(t)={}&
  \bigl(\dot p(t)+\nabla\V_1(q(t))\bigr)
  \cdot\bar v=
  -\nabla\V_0(q(t))\cdot\bar v>0\quad
  \forall t\,,\cr}
  \eqno(6.15)$$
so that $g_0$ is decreasing. It is bounded from
below  because the integral is infinitesimal and
$q(t)\cdot\bar v\ge q_{\sss E}\cdot\bar v$  since
$q(t)\in q_{\sss E}+\D$ (see Lemma~4.7). 

\noindent
Having proved that $g(t)$ is bounded as
$t\to+\infty$, we get in particular that each
$q(t)\cdot v_\alpha$ is bounded for each
$\alpha\in I_0$. We can write
$$g^{\prime\prime}(t)=
  -\sum_{\alpha\in I_0}
  f_\alpha^\prime(q(t)\cdot v_\alpha)\,
  v_\alpha\cdot\bar v-
  \nabla\V_1(q(t))\,.
  \eqno(6.16)$$
The last term is infinitesimal as $t\to+\infty$,
whilst each $-f_\alpha^\prime(q(t)\cdot
v_\alpha)$ is bounded below by the positive
constant 
$$\min\{-f_\alpha^\prime(x)\;:\;
  q_{\sss E}\cdot v_\alpha\le x\le\sup_{t\ge0}
  q(t)\cdot v_\alpha\}
  \eqno(6.17)$$
and $v_\alpha\cdot\bar v>0$ for all
$\alpha\in I_0$. 

\noindent
We get finally
$g^{\prime\prime}(t)\ge\hbox{constant}>0$ as
$t\to+\infty$, which contradicts the boundedness
of~$g$. \qed

      \vfill\eject

%%%%%%%%%%%%%%%%%
\centerline{{\bfuno 8. Global Integrability for
              Finite-sum Potentials}} 
\bigskip

We are going to gather here the statements of
the global integrability results, whose proofs
are scattered in the previous Sections,
concerning the potentials which can be
written as finite sums of one-dimentional
functions (``finite-sum potentials''):
$$\V(q):=\sum_{\alpha=1}^N
  f_\alpha(q\cdot v_\alpha)\,,
  \eqno(8.1)$$
where $v_1,\ldots,v_N$ are given nonzero vectors
in~$\Rn$ ($N\ge1$, no relation to $n$), and the
functions $f_1,\ldots,f_N$ are real functions of
one variable, whose domains are each
either $\R$ or the interval $]0,+\infty[$. The
potential $\V$ is itself defined on the set 
$$\{q\in\Rn\;:\;q\cdot
  v_\alpha\in\dom
  f_\alpha\quad
  \forall\alpha=1,\ldots,N\}\,.
  \eqno(8.2)$$

\bigskip

{\bf Hypotheses 8.1 } \sl 
The vectors
$v_1,\ldots,v_N$ are nonzero and the cone
generated by them is proper.
The $f_\alpha$ are
$C^{m+1}$ ($m\ge2$) functions and
$$\eqalignno{&\sup f_\alpha=+\infty\,,\qquad
  \inf f_\alpha=0\,,&(8.3)\cr
  &f_\alpha^\prime(x)<0\quad\forall x\in
  \dom\,f_\alpha\,,
  &(8.4)\cr
  &f_\alpha^{(k)}(x)\cases{
     >0&if $k$ is even,\cr
     <0&if $k$ is odd,\cr}
                        \quad\forall x\ge a\,,
  &(8.5)\cr
  &f_\alpha^{(m+1)} \hbox{ is monotone on
           $[a,+\infty[$,}
  &(8.6)\cr}$$
where $a\ge0$ is a constant.  
Moreover, whichever one of the three following
conditions i), ii), iii) holds:

\medskip

\item{i)} the vectors $v_\alpha$ verify 
$v_\alpha\cdot v_\beta\ge0$
$\forall\alpha,\beta$;

\smallskip

\item{ii)} all the functions
$f_\alpha$ are multiples of a single
function~$f$:
$$f_\alpha=c_\alpha f\,,\quad c_\alpha>0\,,
  \eqno(8.7)$$
such that 
$$x\mapsto x|f^\prime(x)| 
  \hbox{ is weakly  decreasing on }[a,+\infty[\,;
  \eqno(8.8)$$

\item{iii)} for all
$\alpha=1,\ldots,N$, the function
$f_\alpha$ is such that \rm 
$$\int_a^{+\infty}\mskip-14mu
  f_\alpha(x)\,dx<+\infty\,.
  \eqno(8.9)$$
\rm

\bigskip

{\bf Theorem 8.2 } \sl Assume Hypotheses 8.1.
Then the Hamiltonian system 
$$\dot p=-\nabla\V\,,\quad
  \dot q=p\,,$$
where $\V$ is given by (8.1), is $C^m$-completely
integrable. \rm

     \vfill\eject

%%%%%%%%%%%%%%%%%
\centerline{{\bfuno 9. Appendix}}  
\bigskip

We provide here the detailed proof of a simple
fact needed in Section~3.

\bigskip

{\bf Proposition 9.1 } \sl Let $\{f_i\;:\;i\in
I\}$ be a nonempty family of functions
$\R\to\R$, all of them Lipschitz with the same
constant~$M$:
$$\bigl|f_i(x)-f_i(y)\bigr|\le M|x-y|
  \qquad\forall x,y\in\R\,,\quad
  \forall i\in I\,.
  \eqno(9.1)$$
Let $f(x):=\inf\{f_i(x)\;:\;i\in I\}$ be the
pointwise infimum of the family. If $f$~is
finite at a point $x_0\in\R$, then it is finite
everywhere and Lipschitz with constant~$M$.
\rm

\bigskip

{\bf Proof.} Let $y\in\R$, $i\in I$. Then
$$f_i(y)\ge f_i(x_0)-M|y-x_0|\ge
  f(x_0)-M|y-x_0|\,,
  \eqno(9.2)$$
so that
$$f(y)\ge f(x_0)-M|y-x_0|>-\infty\,.
  \eqno(9.3)$$
The finiteness of~$f$ and half of the Lipschitz 
property are settled ($x_0$ becomes generic
now). Let $\varepsilon>0$ and $i\in I$ such that
$f_i(x_0)\le f(x_0)+\varepsilon$. Then, for
any~$y\in\R$,
$$f_i(y)\le f_i(x_0)+M|y-x_0|\le f(x_0)+
  \varepsilon+M|y-x_0|\,,
  \eqno(9.4)$$
so that
$$f(y)\le f(x_0)+\varepsilon+M|y-x_0|
  \qquad\forall\varepsilon>0\,,
  \eqno(9.5)$$
and the proof is complete. \qed

  \vfill\eject

%%%%%%%%%%%%%%%%%
\centerline{{\bfuno 10. References}} 
\bigskip

\frenchspacing

\item{[A]} Arnold, V. I. (ed.) (1988). 
   {\bf Encyclopaedia of mathematical sciences 3,
     Dyna\-mical Systems III.} 
   Springer Verlag, Berlin.

\medskip

\item{[Ga]} Galperin, G.A. (1982) {\it Asymptotic
   behaviour of particle motion under repulsive
   forces}. {\bf Comm. Math. Phys. 84},
   pp.~547--556.

\medskip

\item{[Gu1]} Gutkin, E. (1985). 
   {\it Integrable  Hamiltonians with
   exponential potentials}. 
   {\bf Physica~D~16},
   pp.~398--404,
   North Holland, Amsterdam.

\medskip

\item{[Gu2]}  Gutkin, E. (1985). 
   {\it Asymptotics of  trajectories 
   for cone potential}. 
   {\bf Physica~D~17},
   pp.~235--242.

\medskip 

\item{[Gu3]}  Gutkin, E. (1987). 
   {\it  Continuity of scattering data 
   for particles on the line 
   with directed repulsive interactions}.
   {\bf J.~Math. Phys. 28}, 
   pp.~351--359.

\medskip

\item{[Gu4]} Gutkin, E. (1988).
  {\it Regularity of scattering trajectories in
  Classical Mechanics}.
  {\bf Comm. Math. Phys. 119},
  pp.~1--12.

\medskip

\item{[GZ1]}  Gorni, G., \& Zampieri, G. (1989). 
  {\it Complete integrability for Hamiltonian
  systems with a cone potential}.  To appear in
  {\bf J. Diff. Equat.}

\medskip

\item{[GZ2]}  Gorni, G., \& Zampieri, G. (1989). 
  {\it Reducing scattering
  problems under  cone potentials  to
  normal form by global canonical
  transformations}. 
  To appear in {\bf J. Diff. Equat.}

\medskip

\item{[He]} Herbst  (1974). {\it Classical
scattering  with long range forces}. \bf Comm.
Math. Phys. 35\rm, pp.~193--214.

\medskip

\item{[Hu]} Hubacher A. (1989). {\it Classical
scattering theory in one dimension}. \bf Comm.
Math. Phys. 123\rm, pp.~353--375.

\medskip

\item{[LL]} Lakshmikantham V., \& Leela, S.
  (1969).
  {\bf Differential and integral inequalities}.
  Volume~I. Academic Press, New York and London.

\medskip 

\item{[MN]} Moauro, V., \& Negrini, P. (1989).
  {\it On the inversion of Lagrange-Dirichlet
  theorem}. {\bf Differ. Integ. Equat.~2},
  pp.~471--478.

\medskip

\item{[MNO]}  Moauro, V., Negrini, P., \& Oliva,
  W.M. (1989). {\it Analytic integrability for a
  class of cone potential mechanical systems}.
  In preparation.

\medskip

\item{[M]} Moser, J. (1983). 
   {\it Various aspects  of integrable
   Hamiltonian systems}. 
   In {\bf Dynamical Systems}
   (C.I.M.E. Lectures, Bressanone 1978),
   pp.~233--290, sec. print.,
   Birkh\"auser, Boston.

\medskip

\item{[OC]} Oliva, W.M., 
   \& Castilla M.S.A.C. (1988).
   {\it On a class of $C^\infty$-integrable 
   Hamiltonian systems}. 
   To appear  in \bf Proc. Royal Society
   Edinburgh\rm.

 \medskip

\item{[S]} Simon B. (1971). {\it Wave operators
 for classical particle scattering}.
\bf Comm. Math. Phys. 23\rm, pp.~37--48.

\bigskip 
\centerline{\hbox to3cm{\hrulefill}}

 \vfill\eject
\end